%% file: main.tex
\documentclass[conference]{IEEEtran}
\IEEEoverridecommandlockouts
\usepackage[x11names,dvipsnames,table]{xcolor}
\def\BibTeX{{\rm B\kern-.05em{\sc i\kern-.025em b}\kern-.08em
    T\kern-.1667em\lower.7ex\hbox{E}\kern-.125emX}}

\input{macros.tex}

\begin{document}

\title{Affinity-aware Serverless Function Scheduling%
\thanks{Paper accepted at 22nd IEEE International Conference on Software Architecture (ICSA 2025). Research partly supported by project PNRR CN HPC - SPOKE 9 - Innovation Grant LEONARDO - TASI - RTMER funded by the NextGenerationEU European initiative through the MUR, Italy (CUP: J33C22001170001),
the research project FREEDA (CUP: I53D23003550006) funded by the framework PRIN 2022 (MUR, Italy), the French ANR project SmartCloud ANR-23-CE25-0012, and the Independent Research Fund Denmark, grant no. 4283-00007B.}
}

\author{\IEEEauthorblockN{Giuseppe De Palma\IEEEauthorrefmark{1}\IEEEauthorrefmark{2}, Saverio Giallorenzo\IEEEauthorrefmark{1}\IEEEauthorrefmark{2}, Jacopo Mauro\IEEEauthorrefmark{3}, Matteo Trentin\IEEEauthorrefmark{1}\IEEEauthorrefmark{2}\IEEEauthorrefmark{3}, Gianluigi Zavattaro\IEEEauthorrefmark{1}\IEEEauthorrefmark{2}
}
\and \IEEEauthorblockA{\IEEEauthorrefmark{1}\textit{Università di Bologna} \\ Bologna, Italy}
\and
\IEEEauthorblockA{\IEEEauthorrefmark{2}\textit{INRIA} \\ Sophia Antipolis, France}
\and
\IEEEauthorblockA{\IEEEauthorrefmark{3}\textit{University of Southern Denmark} \\ Odense, Denmark}
\and
\{giuseppe.depalma2, saverio.giallorenzo2,
matteo.trentin2, gianluigi.zavattaro\}@unibo.it, mauro@imada.sdu.dk}

\maketitle

\begin{abstract}
Functions-as-a-Service (FaaS) is a Serverless Cloud paradigm where a platform
manages the scheduling (e.g., resource allocation, runtime environments) of
stateless functions. Recent work proposed using domain-specific languages to
express per-function policies, e.g., policies that enforce the allocation on
nodes that enjoy lower latencies to databases and services used by the function.
Here, we focus on \emph{affinity-aware} scenarios, i.e., where, for performance
and functional requirements, the allocation of a function depends on the
presence/absence of other functions on nodes.

We present \appp, an extension of a declarative, platform-agnostic language that
captures affinity-aware scheduling at the FaaS level. We implement an
\appp-based prototype on Apache OpenWhisk. Besides proving that a FaaS platform can capture affinity awareness using \appp and improve performance in affinity-aware scenarios, we use our prototype to show that aAPP imposes no noticeable overhead in scenarios without affinity constraints.
\end{abstract}


\input{introduction}

\input{app_syntax}

\input{implementation}

\input{benchmarks.tex}

\input{related_work}

\bibliographystyle{ieeetr}
\bibliography{biblio}

\end{document}

%% file: macros.tex
\usepackage[ruled,vlined]{algorithm2e}
\usepackage[]{graphicx}
\usepackage[]{url}

\usepackage[x11names,dvipsnames,table]{xcolor}
\usepackage{cite}
\usepackage{array}
\usepackage{amsmath}
\usepackage{mathtools}
\usepackage{inconsolata}
\usepackage{amsfonts}
\usepackage{wrapfig}
\usepackage{adjustbox}
\usepackage{mdframed}
\usepackage{nicefrac}

\usepackage{hyperref}
\usepackage[capitalise]{cleveref}

\usepackage{makecell}
\usepackage{xspace}

\usepackage{pgfplots}
\usepackage{listofitems}

\usepackage{tikz}
\usetikzlibrary {
  shapes.geometric,
  arrows.meta,
  bending,
  positioning,
  calc,
  decorations.pathreplacing,
  fit,
  backgrounds
}

\usepackage{pifont}

\usepackage{listings}
\crefname{lstlisting}{listing}{listings}
\Crefname{lstlisting}{Listing}{Listings}

\newcommand\YAMLkeystyle{\color{black}\ttfamily}
\newcommand\YAMLvaluestyle{\color{blue}\ttfamily}

\makeatletter

\newcommand\language@yaml{yaml}

\expandafter\expandafter\expandafter\lstdefinelanguage
\expandafter{\language@yaml}
{
keywords={true,false,null,y,n,fail},
keywordstyle=\ttfamily\color{blue},
basicstyle=\YAMLkeystyle,                                 
sensitive=false,
comment=[l]{\#},
morecomment=[s]{/*}{*/},
commentstyle=\color{purple}\ttfamily,
stringstyle=\YAMLvaluestyle\ttfamily,
moredelim=[l][\color{orange}]{\&},
moredelim=[l][\color{magenta}]{*},
moredelim=**[il][\hlstr{:}\YAMLvaluestyle]{:},   
morestring=[b]',
morestring=[b]",
literate =    {---}{{\ProcessThreeDashes}}3
{>}{{\textcolor{red}\textgreater}}1
{|}{{\textcolor{red}\textbar}}1
{\ -\ }{{\mdseries\ -\ }}3,
}

\lstdefinelanguage{Lisp}{
morekeywords={goal,exists,=},
keywordstyle=\ttfamily\color{RoyalBlue},
literate={=}{{{\color{RoyalBlue}\ttfamily\bfseries =}\color{black}}}{1}
{:}{{{\color{RoyalBlue}\ttfamily\bfseries :}\color{black}}}{1}
{(}{{{\color{gray}\ttfamily\bfseries (}\color{black}}}{1}
        {)}{{{\color{gray}\ttfamily\bfseries )}\color{black}}}{1}
  }

\definecolor{base03}{RGB}{0, 43, 54}
\definecolor{base02}{RGB}{7, 54, 66}
\definecolor{base01}{RGB}{88, 110, 117}
\definecolor{base00}{RGB}{101, 123, 131}
\definecolor{base0}{RGB}{131, 148, 150}
\definecolor{base1}{RGB}{147, 161, 161}
\definecolor{base2}{RGB}{238, 232, 213}
\definecolor{base3}{RGB}{253, 246, 227}
\definecolor{yellow}{RGB}{181, 137, 0}
\definecolor{orange}{RGB}{203, 75, 22}
\definecolor{red}{RGB}{220, 50, 47}
\definecolor{magenta}{RGB}{211, 54, 130}
\definecolor{violet}{RGB}{108, 113, 196}
\definecolor{blue}{RGB}{38, 139, 210}
\definecolor{cyan}{RGB}{42, 161, 152}
\definecolor{green}{RGB}{133, 153, 0}

\lstdefinelanguage{Python}{
  keywords={typeof, null, catch, switch, in, int, str, float, self, throw, import, return, class, if, elif, endif, while, do, else, for, catch, def, raise, not},
  keywordstyle=\color{violet}\ttfamily,
  ndkeywords={True, False},
  ndkeywordstyle=\color{red}\ttfamily,
  identifierstyle=\color{black}\ttfamily,
  sensitive=false,
  comment=[l]{\#},
  morecomment=[s]{/*}{*/},
  commentstyle=\color{base00}\ttfamily,
  stringstyle=\color{green}\ttfamily,
  morestring=[b]',
  morestring=[b]"
}

\newenvironment{codeEnv}[1][base3]
{\begin{mdframed}[backgroundcolor=#1,innertopmargin=-2,innerbottommargin=-2,linecolor=#1!80!black]}
{\end{mdframed}}

\newcommand{\code}[1]{\lstinline[language=yaml]{#1}}

\lst@AddToHook{EveryLine}{\ifx\lst@language\language@yaml\YAMLkeystyle\fi}
\makeatother

\newcommand\ProcessThreeDashes{\llap{\color{cyan}\mdseries-{-}-}}

\usepackage{tikz}

\definecolor{E}{RGB}{51,153,255}

\definecolor{S}{RGB}{6,153,0}

\definecolor{B}{RGB}{153,0,0}

\newcommand{\hln}[1]{{\color{RoyalBlue}\texttt{#1}}}
\newcommand{\hlopt}[1]{{\color{purple}\texttt{#1}}}
\newcommand{\hlstr}[1]{{\color{black}\textbf{\texttt{#1}}}}
\newcommand{\many}[1]{\overline{#1}}
\newcommand{\Div}{\ |\ }

\newcommand{\fopt}{\textit{f\_opt}}
\newcommand{\wopt}{\textit{w\_opt}}
\newcommand{\sopt}{\textit{s\_opt}}
\newcommand{\iopt}{\textit{i\_opt}}
\newcommand{\aopt}{\textit{a\_opt}}

\newcommand{\mOpt}[1]{\colorbox{gray!12}{\(#1\)}}

\newcommand{\app}{\textsf{APP}\xspace}
\newcommand{\appp}{\texorpdfstring{\textsf{\MakeLowercase{a}APP}}{}\xspace}

\newcommand{\smallpar}[1]{\\\indent\textit{#1}.\ }
\Crefname{equation}{Eq.}{Eqs.}
\Crefname{figure}{Fig.}{Figs.}
\Crefname{tabular}{Tab.}{Tabs.}
\Crefname{section}{Sec.}{Secs.}

\usepackage{todonotes}

%% file: introduction.tex
\begin{figure*}[t]
  \center
  \noindent\begin{minipage}{.8\textwidth}
    \includegraphics[width=\textwidth]{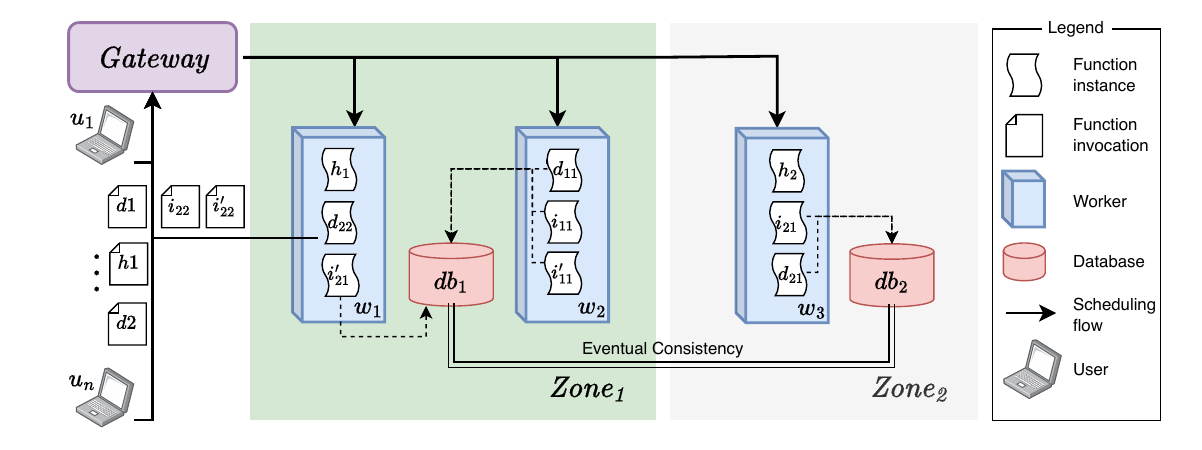}
  \end{minipage}
\begin{minipage}{.18\textwidth}
\begin{codeEnv}
\begin{lstlisting}[xleftmargin=-.8em, language=yaml, mathescape, numbers=left, numberstyle=\tiny,basicstyle=\linespread{0.8}\ttfamily]
- d:
  - $\hln{workers}\hlstr{:}\ \hlopt{*}$
    $\hln{affinity}$: 
      - !h
- i:
  - $\hln{workers}\hlstr{:}\ \hlopt{*}$
    $\hln{affinity}$: 
      - !h
      - d
- h:
  - $\hln{workers}\hlstr{:}\ \hlopt{*}$
    $\hln{affinity}$: 
      - !d
      - !i
\end{lstlisting}
\end{codeEnv}
\end{minipage}
\caption{Examples of a FaaS infrastructure (left) and an \appp script (right).}
  \label{fig:example}
  \end{figure*}

\section{Introduction}
\label{sec:intro}

Functions-as-a-Service (FaaS) is a programming paradigm supported by the Serverless Cloud
execution model~\cite{Jonas-etal:BerkeleyViewOnServerless}. In FaaS,
developers implement a distributed architecture from the composition of
stateless functions and delegate concerns like execution runtimes and resource allocation to the serverless platform, thus focusing on writing code that implements business logic rather than worrying about infrastructure management.
The main cloud
providers offer FaaS~\cite{web:IntroducingAwsLambda,web:googlefunctions,web:azurefunctions} and open-source alternatives exist too~\cite{web:OpenWhisk,web:openfaas,HSHVAA16,web:fission}.

A common denominator of these platforms is that they manage the allocation of
functions over the available computing resources, also called \emph{workers},
following opinionated policies that favour some performance principle.
Indeed, effects like \emph{code locality}~\cite{HSHVAA16}---due to latencies in
loading function code and runtimes---or \emph{session
locality}~\cite{HSHVAA16}---due to the need to authenticate and open new
sessions to interact with other services---can substantially increase the run time of
functions.
The breadth of the design space of serverless scheduling policies is
witnessed by the growing literature focused on techniques that mix one or more
of these locality principles to increase the performance of function
execution, assuming some locality-bound traits of functions~\cite{CAVJLLPPR20,KGB20,BS21,ZE21,SAVA21,SJCGB22}.
Besides performance, functions can have functional requirements that the scheduler shall consider. For example, users might want to ward off
allocating their functions alongside ``untrusted'' ones---common threat vectors
in serverless are limited function isolation and the ability of functions to
(surreptitiously) gather weaponisable information on the runtime, the
infrastructure, and the other
tenants~\cite{BCCCFIMMRS17,WLZRS18,AFFRSSW18,PPTMAA20}.

Although one can mix different principles to expand the profile coverage of a
given platform-wide scheduler policy, the latter hardly suits all kinds of
scenarios. This shortcoming motivated De Palma et al.~\cite{PGMZ20,DGMTZ22} to
introduce a domain-specific, platform-agnostic, declarative language, called
\emph{Allocation Priority Policies} (\app) to specify custom function allocation
policies.
Thanks to \app, the same platform can support different scheduling policies, each tailored to meet the specific needs of a set of related functions.
%
%
De Palma et al.\@ validated their approach by implementing an \app-based
serverless platform as an extension of the open-source Apache OpenWhisk project~\cite{PGMZ20,DGMTZ24}.

Our contributions originate from the observation that, at lower levels of the
cloud stack, popular Infrastructure-as-a-Service (IaaS) platforms (e.g.,
OpenStack~\cite{web:openstack_affinity}) and Container-as-a-Service (CaaS)
systems (e.g., Kubernetes~\cite{web:kube_affinity}) allow users to express
affinity and anti-affinity constraints about the
allocation of VM/containers---e.g., reducing overhead, by shortening data paths
via co-location (affinity), increasing reliability, by evenly distributing
VM/containers among different nodes (anti-affinity) or improving security, by
preventing the co-location of VM/containers belonging to different trust tiers
(anti-affinity).
On the contrary, FaaS platforms do not natively support the possibility to
express affinity-aware scheduling,
where function allocation depends on the presence (affinity) or absence
(anti-affinity) at scheduling time of other functions in execution on the
available workers.
%

\paragraph*{Contribution}
Recognising the potential of FaaS-level affinity-aware scheduling policies, we
propose a language-based solution, obtained by extending \app into an
affinity-aware function scheduling language called \appp. In
\cref{sec:IntroExample}, we present an example of affinity-aware scheduling at
the FaaS level that we use to informally introduce \appp. We formalise our
proposal in~\cref{sec:app}, where we present the \appp syntax and discuss the
increment of expressiveness w.r.t.~\app. 
In \cref{sec:implementation}, we concretise our proposal by presenting a
prototype implementation of an \appp-based serverless platform, namely, an
extension of Apache OpenWhisk able to enforce \appp-defined FaaS (anti-)affinity
scheduling constraints. In \cref{sec:affinity_aware}, we experimentally show
that the usage of (anti-)affinity constraints are beneficial by considering an
implementation of the affinity-aware scenario introduced in
\cref{sec:IntroExample}. In \cref{sec:non_affinity_aware}, we compare the
performance of our \appp-based prototype and vanilla OpenWhisk with 7 benchmarks
to show that \appp imposes negligible overhead. We discuss related work and draw
concluding remarks in \cref{sec:conclusion}.

\begin{figure*}[t]
  \begin{minipage}{.66\textwidth}
  \begin{adjustbox}{width=1.05\textwidth}
  \begin{minipage}{\textwidth}
  \vspace{-1.5em}
  \[
  \begin{array}{lll}
  && \hspace{.12\textwidth} \textit{id} \in \textit{Identifiers} 
  \hspace{.1\textwidth} n \ \in \ \mathbb{N} 
  \\[.5em]
  \textit{app}         & \Coloneqq & \many{- \textit{tag}}
  \\
  \textit{tag}         & \Coloneqq & \textit{id} \ \hlstr{:} \ \many{\mathtt{-}\ \textit{block} } \quad \mOpt{\hln{followup}\ \hlstr{:}\ \fopt}
  \\
  \textit{block}       & \Coloneqq & \hln{workers} \ \hlstr{:}\ \wopt
  \quad \mOpt{\hln{strategy} \ \hlstr{:}\ \sopt}\\&&
  \quad \mOpt{\hln{invalidate} \ \hlstr{:}\ \many{-\ \iopt}}
  \quad \mOpt{\hln{affinity} \ \hlstr{:}\ \many{-\ \aopt}}
  \\
  \wopt      & \Coloneqq & \hlopt{*} \Div \many{\mathtt{-}\ \textit{id}}
  \\
  \sopt      & \Coloneqq & \hlopt{any} 
                \Div \hlopt{best\_first}
  \\
  \iopt      & \Coloneqq & \hlopt{capacity\_used} \ n \hlopt{\%} \Div \hlopt{max\_concurrent\_invocations} \ n 
  \\
  \aopt & \Coloneqq & \textit{id} \Div \textit{!id}
  \\
  \fopt      & \Coloneqq & \hlopt{default} \Div \hlopt{fail}
  \end{array}
  \]
  \end{minipage}
  \end{adjustbox}
  \caption{\label{fig:app_syntax} \appp syntax.}
  \end{minipage}
  \hspace{-.5em}
  \begin{minipage}{.3\textwidth}
  \begin{adjustbox}{width=\textwidth}
  \begin{minipage}{.8\textwidth}
  \vspace{.5em}
  \begin{codeEnv}
  \begin{lstlisting}[xleftmargin=-1.1em, language=yaml, mathescape, basicstyle=\linespread{1}\footnotesize\ttfamily,numbers=none]
  - f_tag:
    - $\hln{workers}$:
      - local_w1
      - local_w2
      $\hln{strategy}\hlstr{:}\ \hlopt{best\_first}$
      $\hln{invalidate}$:
      - $\hlopt{capacity\_used}$ 80%
      $\hln{affinity}\hlstr{:}$ g_tag,!h_tag
    - $\hln{workers}$:
      - public_w1
    $\hln{followup}\hlstr{:}\ \hlopt{fail}$
  \end{lstlisting}
  \end{codeEnv}
  \end{minipage}
  \end{adjustbox}
  \caption{\label{fig:app_example} Example \appp script.}
  \end{minipage}
  \end{figure*}

\section{Example of an Affinity-aware FaaS Scenario}\label{sec:IntroExample} We have a
\emph{divide-et-impera} data-crunching serverless application implemented
through two companion functions. The first, invoked by the users, is called
\textit{divide}. Its task is to split some data into chunks, store them in a
database, and invoke instances of the second function. The second function,
invoked by the \textit{divide} for each stored chunk, is called
\textit{impera}. Its task is to retrieve and process a chunk of data from the database.

We run the above functions on the FaaS infrastructure depicted on the left of
\cref{fig:example}. The infrastructure includes two zones (e.g., separate
regions of a cloud provider) and it has a \(\textit{Gateway}\) that decides on
which worker to allocate the execution of the functions. The infrastructure also includes three
workers:
\(w_1\) and \(w_2\) in \(\textit{Zone}_1\) and \(w_3\) in \(\textit{Zone}_2\).
Each zone hosts an instance of an eventually-consistent distributed
database~\cite{V09}, used by the functions running in that
zone---eventually-consistent systems are typical for (FaaS)
scenarios like ours, where one favours throughput and availability
w.r.t.\ e.g., overall data consistency~\cite{BG13}.

In \cref{fig:example}, we represent function allocation requests with labelled
document icons sent to the \(\textit{Gateway}\). Note that the users (the
laptop icons in \cref{fig:example}) launch the \textit{divide} function (e.g.,
\(d_3\)) while the running \textit{divide} (e.g., \(d_2\) requesting
\(i_2\) and \(i'_2\)) invoke the \textit{impera} functions.

Our FaaS infrastructure executes other functions besides the one above.
In \cref{fig:example}, we represent these requests with the labels \(h_{1}\),
\(h_{2}\), and \(h_{3}\) which are compute-intensive functions---called
\textit{heavy}---that use a high amount of computational resources of the worker
running them.

Given this context, an initial example of an affinity-aware scheduling policy is
to avoid the co-occurrence of the \textit{divide} and \textit{impera} functions
with the \textit{heavy} ones. In this way, we can improve the performance of
\textit{divide} and \textit{impera} by avoiding resource contention with the
\textit{heavy} functions.
Another improvement regards the interaction with the database. The
eventually-consistent behaviour of the database entails possible delays to
synchronise the instances. 
Waiting for synchronisation is necessary only when the functions accessing the
database connect to
different database instances.
Moreover, to further reduce delay, we can exploit the principle of \emph{session
locality} and let functions running on the same worker share the same connection
with the database. This affinity-aware scheduling policy places \textit{impera}
functions only on workers that already host \textit{divide} functions and avoid
the overhead of re-establishing new connections.

%
%



%
These constraints can be encoded in \appp as shown in the script in \cref{fig:example}.
In the code, we find three top-level items: \code{d}, \code{i}, and \code{h},
which are tags that identify policies, each describing the
scheduling logic of a set of related functions.
In the example, the tag \code{d} describes the logic for the \textit{divide}
functions while \code{i} and \code{h} target respectively the \textit{impera}
and \textit{heavy} ones. 
The line \hln{workers}\hlstr{:}\ \hlopt{*} found under
all tags indicates that their related functions can use any of the available
workers. 
From the top, 
under tag \code{d}, we use the \hln{affinity} clause, introduced by \appp, to
specify that \code{d}-tagged functions should \emph{not} be scheduled 
on a worker that currently hosts \textit{heavy} functions (\code{!h}).
Specifically, this is an example of \emph{anti-affinity}, where we prevent the
allocation of the tagged functions (e.g., \code{d}) on a worker that already
hosts any anti-affine function (e.g., tagged \code{h}). Tag \code{i} declares the same
anti-affinity for \textit{heavy} functions, but it also indicates that
\code{i}-tagged functions are \emph{affine} with \code{d}-tagged ones. Affinity
means that we can schedule a function on a candidate worker only if it currently
hosts the former's affine functions. In the example, we use affinity to have
\textit{impera} functions run in the same worker of \textit{divide} functions.
Finally, we use tag \code{h} to complement the anti-affinity relation expressed
in the previous tags, i.e., the \textit{heavy} functions are anti-affine with
both \code{d} and \code{i} functions and shall not be scheduled in workers that
already host any of the latter.

Notably, we purposefully do not identify who writes the \appp script in the
example, e.g., the developer of the functions or the administrator of the
platform. Indeed, \appp (in general, \app and all its extensions) caters to
different cloud stakeholders for scheduling policy definition. For instance, if
we contextualise our example in a local private cloud setup, then users can directly
write their own \appp scripts because they have direct knowledge of the
infrastructure nodes. Contrarily, if we are in a managed cloud environment, the
cloud provider would use \appp to implement and enforce scheduling requirements
specified by their clients based on their workflows---e.g., synthesising \appp
scripts from function and workflow code~\cite{DGLMTZ23,DGMTZ24c}.

%% file: app_syntax.tex
\section{The \appp Language}
\label{sec:app}

We now present \appp, our extension of the
FaaS function scheduling language \app~\cite{PGMZ20,DGMTZ22} 
with affinity and anti-affinity constraints. 

We report in \cref{fig:app_syntax} the syntax of \appp. From here on, we
indicate syntactic units in \emph{italics}, optional fragments in
\(\mOpt{grey}\), terminals in \code{monospace}, and lists with \(\many{bars}\).
The syntax of \appp draws inspiration from YAML ~\cite{web:YAML}, a renowned
data-serialisation language for configuration files---e.g., many modern
cloud tools, like Kubernetes and Ansible, use this format.\footnote{While \appp
scripts are YAML-compliant, for presentation, we stylise the syntax to increase
readability. For instance, we omit quotes around strings, e.g., \hlopt{*}
instead of \hlopt{"*"}.}
In \appp, functions have associated a tag that identifies some scheduling
policies. An \appp script represents: \textit{i)} named scheduling policies
identified by a \textit{tag} and \textit{ii)} policy \(\textit{block}\)s that
indicate either some collection of workers, each identified by a worker
\textit{id}, or the universal \hlopt{*}. To schedule a function, we use its tag
to retrieve the scheduling policy that includes one or more blocks of possible
workers. To select the worker, we iterate top-to-bottom on the blocks. We stop
at the first block that has a non-empty list of valid workers and then select
one of those workers according to the strategy defined by the block (described
later).

\begin{figure*}[t]
 \begin{tikzpicture}[
   component/.style={rectangle, draw, rounded corners, minimum width=3cm, minimum height=1cm, font=\bfseries, align=center},
   modified/.style={rectangle, draw, rounded corners, fill=blue!30, minimum width=2.5cm, minimum height=1cm, font=\bfseries, align=center},
   new/.style={rectangle, draw, rounded corners, fill=yellow!30, minimum width=2.5cm, minimum height=1cm, font=\bfseries, align=center},
   arrow/.style={->, thick},
   every node/.style={font=\sffamily}
]

\node[component] (loadbalancer) {Entrypoint\\(Nginx)};
\node[component, right=1cm of loadbalancer, minimum width=5cm, minimum height=5.5cm, align=center] (controller) {};

\node[anchor=north, font=\bfseries, yshift=-0.3cm] at (controller.north) {Controller};

\node[modified, below=0.5cm of controller.north, yshift=-0.4cm, minimum height=0.6cm] (aappparser) {Parser};
\node[component, below=0.2cm of aappparser, minimum width=4.5cm, minimum height=3.5cm, align=center] (configurableLB) {};
\node[anchor=north, font=\bfseries, yshift=-0.3cm] at (configurableLB.north) (configurableLBLabel) {ConfigurableLoadBalancer};

\node[modified, below=0.2cm of configurableLBLabel, minimum height=0.6cm] (scheduler) {Schedule function};
\node[new, below=0.2cm of scheduler, minimum height=0.6cm] (activefunctions) {activeFunctions Table};
\node[new, below=0.2cm of activefunctions, minimum height=0.6cm] (activetagactivations) {activeTagActivations Table};

\node[component, right=1cm of controller] (kafka) {Message Broker\\(Kafka)};
\node[component, right=1cm of kafka] (invoker) {Invoker\(_{...}\)};
\node[component, above=0.5cm of invoker] (invoker1) {Invoker\(_1\)};
\node[component, below=0.5cm of invoker] (invokern) {Invoker\(_n\)};

\draw[arrow] (loadbalancer) -- (controller);
\draw[arrow] (controller) -- (kafka);
\draw[arrow] (kafka) -- (invoker1);
\draw[arrow] (kafka) -- (invokern);
\draw[arrow] (kafka) -- (invoker);

\end{tikzpicture} 
\caption{Extended Apache OpenWhisk for \appp (modified modules in blue, added modules in yellow).}
\label{fig:openwhisk_extension}
\end{figure*}

Each tag can define a \hln{followup} clause, which specifies what to do if the
policy of the tag did not lead to the scheduling of the function; either
\hlopt{fail}, to terminate the scheduling, or \hlopt{default}, to apply the
special \texttt{default}-tagged policy.
Each block can define a \hln{strategy} for worker selection ($\hlopt{any}$
selects non-deterministically one of the available workers in the list;
$\hlopt{best\_first}$ selects the first available worker in the list), a list of
constraints that \hln{invalidate}s a worker for the allocation
($\hlopt{capacity\_used}$ invalidates a worker if its resource occupation
reaches the set threshold; $\hlopt{max\_concurrent\_invocations}$ invalidates a
worker if it hosts more than the specified number of functions), and an
\hln{affinity} clause that carries a list containing affine tag identifiers
$\textit{id}$ and anti-affine tags, represented by negated tag identifiers
$\textit{!id}$. \appp is a minimal extension of \app where we add the
\hln{affinity} clause to capture (anti-)affinity constrains. Similar to the
notion of affinity introduced by Microsoft in its IaaS
offering~\cite{service_fabric_affinity}, in \appp, the relation of
(anti-)affinity is ``directional'', i.e., we impose no well-formedness property
like symmetry or anti-symmetry on affinity or anti-affinity
to avoid limiting \appp's ability to capture meaningful scenarios.\footnote{For
instance, if we had 
symmetric anti-affinity, we would not capture scenarios where, e.g., a function
\code{init} is the seeding function for a database and function \code{query}
manipulates that data. The function \code{init} should always run before
\code{query} but never where \code{query} is already running, while function
\code{query} should run where \code{init} is present. To obtain this behaviour,
we need \code{init} anti-affine with \code{query} but \code{query} affine with
\code{init}.}

We practically illustrate \appp with the two-block \appp policy example
(for functions tagged \texttt{f\_tag}) found in \cref{fig:app_example}. The
first block restricts allocations on the workers labelled
\lstinline[language=yaml]{local_w1} and \lstinline[language=yaml]{local_w2} and
the latter on \lstinline[language=yaml]{public_w1}. The first block specifies as
invalid (i.e., such that they cannot host the function under scheduling) the
workers that reach a memory consumption above 80\%. Since the strategy is
$\hlopt{best\_first}$, we allocate the function on the first valid worker; if
none are valid, we proceed with the next block. The function has affinity with
\texttt{g\_tag} and anti-affinity with \texttt{h\_tag}. Hence, a valid worker
must host at least a function with tag \texttt{g\_tag} and no
functions with tag \texttt{h\_tag}. 
If both the first and second blocks do not find a valid worker, the scheduling
of the function \hlopt{fail}s (instead of continuing with the \texttt{default}
tag).

Notably, the addition of (anti-)affinity constraints 
increases the
expressiveness of \app. 
Indeed, \app can capture anti-affinity
constraints 
by severely limiting the flexibility of resource allocation,
e.g., either through a partition of the workers and a static assignment
of anti-affine functions to distinct partitions, or by limiting
artificially the capacity that can be used or the number of functions we can 
allocate on a worker. This approach conflicts with the cloud principle of 
resource sharing and optimization. 
The situation for affinity is even poorer: \app cannot capture these constraints
because it does not keep track of the functions allocated on workers.

%% file: implementation.tex
\section{\appp-based Apache OpenWhisk}
\label{sec:implementation}

To validate \appp, we implemented an \appp-based FaaS platform, obtained by the
existing \app-based version of Apache
OpenWhisk.

We support the description of our implementation with \cref{fig:openwhisk_extension}, which illustrates the main components and the typical flow of function invocations in OpenWhisk.
Specifically, the \textit{Entrypoint} for function execution requests is an Nginx reverse proxy, which forwards the requests to a \textit{Controller}. The \textit{Controller}, in turn, routes them via a Kafka \textit{Message Broker} to the \textit{Invoker}s—the workers in OpenWhisk terminology.

In the figure, we highlight 
the main interventions we performed to make the existing \app{}-based OpenWhisk
version
\appp{}-compliant. 
The extension concerns two parts: the \app{} parser and the
\textit{ConfigurableLoadBalancer}, both absent in vanilla OpenWhisk and
originally introduced in \app{}-based OpenWhisk~\cite{PGMZ20}.
We respectively extended the parser and the \textit{ConfigurableLoadBalancer} to
add compatibility for \appp{} scripts and to keep track of the functions
allocated to all the workers.
In particular, we introduced two lookup tables, called \textit{activeFunctions}
and \textit{activeTagActivations}, to implement the tracking functionality. The
first table associates the allocated functions (and their tags) to their host
worker and allows the load balancer to verify (anti-)affinity constraints. The
\textit{activeTagActivations} table keeps tracks of the state of the different
function instances (possibly of the same function definition, so we cannot use
their identifiers) by pairing their \emph{activation} \emph{id}s with their
function identifiers. When we observe the termination of an active function, we
look its function identifier up and remove that instance from the
\textit{activeFunctions} table---we detect instance terminations thanks to the
messages workers send to notify the load balancer of their completion.
Code-wise, these changes required the modification of the \textit{schedule} function of the \textit{ConfigurableLoadBalancer}.


\begin{figure*}[tb]
\begin{lstlisting}[
  language=Python,
  numberblanklines=false,
  label=lst:code,
  basicstyle=\normalsize,
  caption={The pseudo-code of the \code{schedule} and \code{valid} functions.}
]
def schedule(f, conf, aapp, reg):
 (memory, tag) = reg[f]
 blocks = aapp[tag].blocks # get the blocks
 if aapp[tag].followup != 'fail':
  blocks += aapp['default'].blocks # add default tag blocks
 for block in blocks:
  if '*' in block['workers']: 
   block['workers'] = conf.keys
  workers = [ for worker in block['workers'] if valid(f,worker,conf,reg,block)]
  if len(workers) > 0: # if at least one valid worker is found
   if block['strategy'] == 'best_first': 
    return workers[0]
   elif block['strategy'] == 'any': 
    return random.choice(workers)
 raise Exception('Function not schedulable')

def valid(f, w, conf, reg, block):
 (memory, tag) = reg[f]
 if (w not in conf) or (conf[w]['memory_used'] + memory > conf[w]['max_memory']):
  return False
 if 'invalidate' in block:
  if ('capacity_used' in block['invalidate']) and
     (block['invalidate']['capacity_used'] <= conf[w]['memory_used']):
    return False
  if ('max_concurrent_invocations' in block['invalidate']) and
    (block['invalidate']['max_concurrent_invocations'] <= len(conf[w]['fs'])):
   return False
 if 'affinity' in block:
  affine_tags = set([t for t in block['affinity'] if not t.startswith('!')])
  anti_affine_tags = set([t[1:] for t in block['affinity'] if t.startswith('!')])
  w_tags = set([t for (_, t) in [reg(f) for f in conf[w]['fs']]])
  for t in affine_tags:
   if t not in w_tags: return False
  for t in anti_affine_tags:
   if t in w_tags: return False
 return True
\end{lstlisting}
\end{figure*}


The scheduling semantics of \appp scripts is straightforward. We present it in
(Python-like) pseudo-code in \Cref{lst:code}. In \Cref{lst:code}, the
\texttt{schedule} function requires the name of the function under scheduling
(\texttt{f}), a map that represents the current infrastructure configuration
(workers and functions running therein, explained later) (\texttt{conf}), an
\appp script encoded as a Python dictionary of objects (\texttt{aapp}), and a
registry that maps each function to a tag and its memory occupancy
(\texttt{reg}). The infrastructure configuration maps, for each worker, the list
of functions scheduled on it (\texttt{fs}), the memory allocated for those
functions (\texttt{memory\_used}), and the total amount of memory of the worker
(\texttt{max\_memory}).


Given these inputs, \texttt{schedule} gets the tag associated with \texttt{f}
(Line 2) and then extracts the blocks associated with this tag in the
\texttt{aapp} script (Line 3). If the \hln{followup} option is different from
``fail'', we append the blocks associated with the \texttt{default} tag to
the list of \texttt{f}'s blocks (Line 5). Then, we obtain the list of valid
workers for every block in order of appearance (Line 9). When the \hln{workers}
clause uses \hlopt{*}, we consider all the workers present in the configuration
(Line 8). If the list of valid workers is non-empty, we choose the first one
when the strategy is \hlopt{best\_first} (Line 12) and a random one otherwise
(Line 14). If the list is empty, the scheduling fails (Line 15).

The \code{schedule} function uses the \code{valid} function to check when a
worker is valid, i.e., it is available, it has enough capacity to host the
function (Lines 18--19), and that allocating on it the function satisfies all
the constraints of \hlopt{capacity\_used}, \hlopt{max\_concurr\-ent\_in\-voca\-tions}
(Lines 21--26), and \hln{affinity} (Lines 27--34).
%

To implement our prototype, we have modified the Scala codebase of the OpenWhisk
project; specifically, we have modified the scheduling algorithm to implement the
logic of \Cref{lst:code} and manage the workers-functions status, on a fork of
the OpenWhisk repository~\cite{repository}. The entire system is easily
deployable using Terraform and Ansible scripts.

\begin{figure}[t]

\begin{codeEnv}
  \begin{lstlisting}[ xleftmargin=-.8em, language=yaml, mathescape, numbers=left, numberstyle=\tiny,basicstyle=\linespread{0.8}\ttfamily]
  - d:
    - $\hln{workers}\hlstr{:}\ \hlopt{*}$
      $\hln{strategy}\hlstr{:}\ \hlopt{random}$
      $\hln{affinity}$: 
        - !h_eu
        - !h_us

  - i:
    - $\hln{workers}\hlstr{:}\ \hlopt{*}$
      $\hln{strategy}\hlstr{:}\ \hlopt{random}$
      $\hln{affinity}$: 
        - !h_eu
        - !h_us
        - d

  - h_eu:
    - $\hln{workers}$:
        workereu1
  - h_us:
    - $\hln{workers}$:
        workerus1
  \end{lstlisting}
  \end{codeEnv}
  \caption{The \appp script used for the tests.}
  \label{fig:aapp_script}
\end{figure}

\section{Performance Improvements\\ via Affinity-awareness}
\label{sec:affinity_aware}
To validate our platform and show the benefit of \mbox{(anti-)}affinity
constraints for affinity-aware scenarios,
we use the example from \cref{sec:IntroExample} as a benchmark case
study and show that, by enforcing (anti-)affinity constraints, we can reduce
average execution times and tail latency.

Recalling the example, 
we consider 
a simple \emph{divide-et-impera} serverless architecture running in a realistic
co-tenancy context.
Users invoke \emph{divide} functions, requesting the solution of a problem. At
invocation, \emph{divide} splits the problem into sub-problems and invokes
instances of the second function, \emph{impera}. The \emph{impera} instances
solve their relative sub-problems and store their solution fragments on a
persistent storage service. After the \emph{impera}s terminated, \emph{divide}
retrieves the partial solutions, assembles them, and returns the response to the
user. We consider a multi-zone execution context where each zone hosts an
instance of an eventually-consistent distributed database. The workers in one
zone access the local instance of the database. Another function, called
\emph{heavy}, represents possible interferences of serverless co-tenancy.
\begin{figure*}[t]
  \begin{center}
    \begin{tabular}{l | c | c | c}
      \hline
      \textbf{Configuration} & \textbf{Mean Latency (ms)} & \textbf{Median
      Latency (ms)} & \textbf{95\(^{th}\) Tail Latency (ms)} \\
      \hline
      \appp{} & 1547 & 883 & 3041 \\
      anti-affinity-only \appp{} & 2337 (+40\%) & 2381 (+91\%) & 3476 (+13\%) \\
      \app & 8118 (+135\%) & 2648 (+99\%) & 60157 (+180\%) \\
      \hline
    \end{tabular}
    \includegraphics[width=.85\textwidth]{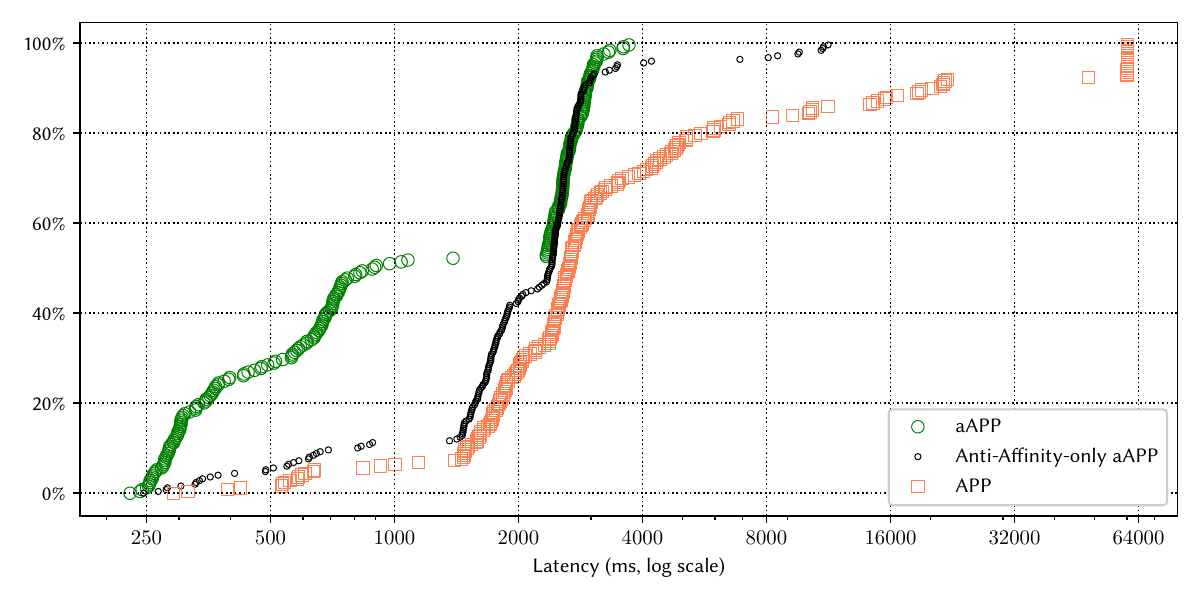}
  \end{center}
  \caption{Top, table reporting the mean, median, and 95\(^{th}\) tail latencies
  of the \textit{divide} functions under \appp{}, anti-affinity-only \appp{},
  and \app (percentages represent variation w.r.t.~\appp{}). Bottom, the sorted
  scatter plot of the latencies incrementally sorted.}
  \label{fig:benchmark}
\end{figure*}
%
%
%
\smallpar{Experimental setup}
To run the case study, we deploy the OpenWhisk versions of \app{} and \appp{} on
an 8-node Kubernetes cluster on the Digital Ocean platform. As schematised in
\cref{table:workers}, one node acts as the control plane (and as such, it is
unavailable to OpenWhisk), one hosts the OpenWhisk core components (i.e., the
Controller, the OpenWhisk internal database CouchDB, and the messaging system
Kafka), 
and six nodes are workers. We deploy the control plane and the OpenWhisk core
components on virtual machines with 2 vCPU and 2 GB RAM, while we deploy 4
workers on virtual machines with 2 vCPU and 2 GB RAM and 2 workers with 1 vCPU
and 1 GB RAM. All machines run the Ubuntu Server 20.04 OS. Location-wise, we
place the control plane, the OpenWhisk core components, and 3 workers in Europe
and 3 workers in North America (2 with the more powerful configuration and 1
with the lesser one in each zone). To implement persistent storage, we deploy a
2-node MongoDB replica set, one in Europe and one in North America, using the
6.0.2 version of the Community Server.

\begin{figure}
  \center
  \begin{tabular}{| c | c | c | c |}
    \hline
    \textbf{Role} & \textbf{vCPU} & \textbf{RAM} & \textbf{Location} \\
    \hline
    \hline
    Control Plane & 2 & 2 GB & Europe \\
    \hline
    OW Core & 2 & 2 GB & Europe \\
    \hline
    OW Worker & 2 & 2 GB & Europe \\
    \hline
    OW Worker & 2 & 2 GB & Europe \\
    \hline
    OW Worker & 2 & 2 GB & North America \\
    \hline
    OW Worker & 2 & 2 GB & North America \\
    \hline
    OW Worker & 1 & 1 GB & Europe \\
    \hline
    OW Worker & 1 & 1 GB & North America \\
    \hline
    MongoDB Replica & 1 & 1 GB & Europe \\ 
    \hline
    MongoDB Replica & 1 & 1 GB & North America \\
    \hline
  \end{tabular}
  \caption{Nodes used for the experimental evaluation.}
  \label{table:workers}
\end{figure}

We distribute the load generated by the \emph{heavy} functions on the platform
with two variants, \textit{heavy\_eu} and \textit{heavy\_us}, which
we constrain to be resp.\@ allocated in the Europe and the North America data
centres on the less powerful workers (identified with \textit{workereu1} and
\textit{workerus1}), 
to further amplify the effect of co-tenancy they exert.

%
%
%
All functions are in JavaScript and run on OpenWhisk NodeJS runtime
\textit{nodejs:14}. The \emph{divide} function invokes two instances of the
\emph{impera} functions with 3 parameters: i) a freshly generated index, ii) an
array populated with 100 random numbers, and iii) the initial--final indexes of
the values to work on (0--49 to the first \emph{impera} instance, 50--99 to the
second one). The \emph{divide} function waits for the \emph{impera} functions to
terminate, and then opens a connection to the local instance of MongoDB with the
aim of retrieving documents representing the results computed by the
\emph{impera} functions. The \emph{divide} function detects the correct
documents by issuing a query on the MongoDB with the freshly generated index.
Each \emph{impera} function simulates a computation on the values received from
the \emph{divide} function by connecting to the local instance of MongoDB and
storing on it one document for each received value. Each of these documents
contains one of the values and the freshly generated index received upon
invocation. In this way, each execution of the \emph{impera} function stores in
MongoDB 50 documents. The \emph{heavy} function simulates computing-resource
consumption by performing 1 billion iterations of a computation consisting of
the random generation of two numbers and the execution of their multiplication.

Note that the \emph{divide} function needs to retrieve from MongoDB the
documents generated by the two \emph{impera} function it invokes. Each function
opens a connection to the local instance of MongoDB and, in case an
\emph{impera} function runs on a different zone w.r.t.\@ the \emph{divide}
function, it can take some time for the two database instances to eventually
become consistent (converge). The \emph{divide} function implements an
exponential back-off retry approach~\cite{osman2018microservices}---it tries to
fetch the documents from its local storage instance; if the data is not there,
starting from a 1-second delay, the function waits for a back-off time that
exponentially increases at each retry.
\paragraph*{Experiments and Results}%
In our experiment, we consider three \app{}/\appp{} scripts to showcase the
benefits of \mbox{(anti-)}affinity constraints.
The first, which uses the full expressiveness of \appp{}, is the one reported in
\cref{fig:aapp_script}---where \emph{impera}s (tagged with \texttt{i}) are
affine with \emph{divide} (tagged with \texttt{d}) and they are both anti-affine
with the \emph{heavy} functions (tagged with \texttt{h\_eu} and \texttt{h\_us}).
We impose affinity between \emph{divide} and \emph{impera} functions to
guarantee that the database writer (the \emph{impera} function) and reader (the
\emph{divide} function) access the same database instance. We impose
anti-affinity with the \emph{heavy} functions to avoid resource contention
between the \emph{divide}/\emph{impera} functions and the \emph{heavy}
functions.
The second script removes the affinity constraints between \emph{impera} and
\emph{divide} from the first script (anti-affinity-only-\appp).
The third script omits the anti-affinity constraints from the second one,
effectively making it an \app{} script.

Each experiment involves 5 sequential runs. 
Each run invokes the \emph{heavy\_eu} and \emph{heavy\_us} functions in
non-blocking mode,
followed by 10 calls of the \textit{divide} function, each one waiting for the
previous to complete.
Upon termination of the \emph{heavy} functions, we proceed
with the remaining runs, for a total of 10 \emph{heavy} and 50 \textit{divide}
functions 
per experiment.
To ensure reliable results, we run the experiment 5 times, totalling 250 calls
of the \textit{divide} function for each of the three \app{}/\appp{} script.
We use Apache JMeter to simulate each request, tracking its latency, number of
retries (to retrieve storage data), and outcomes (success or failure).

We report the mean, median, and 95\(^{th}\) tail latencies for the divide
functions at the top of \cref{fig:benchmark} under \appp{}, Anti-Affinity-Only
\appp{}, and \app{}. From the values, we mainly note that \appp{} has the best
performances, while the latencies increase for both anti-affinity-only \appp{}
and, even more substantially, for \app.

To further analyse the differences between the performance of the three
policies, at the bottom of \cref{fig:benchmark}, we report the sorted scatter
plot of the latencies of the \textit{divide} functions from the shortest to the
longest (\(x\)-axis). We focus on this measure because it offers a comprehensive
overview of the performance of the architecture. In particular, this measure
includes the latencies of the related \textit{impera} functions, whose run times
concretely impact the request-response delay experienced by the users
interacting with the system.

%
The first striking observation we gather is that there is a gap (there are
almost no instances) in the distribution of the \appp data points between the
1000ms and the 2400ms mark. We conjecture that this behaviour derives from
having OpenWhisk core components installed in one region, which exert some
overhead on the workers of the other region when they interact with the platform
(e.g., to fetch functions and receive/send requests/notifications). We see
similar intervals, although less apparent, for \app and
anti-affinity-only-\appp{}.

In the 200--1000ms interval, \appp provides consistent, fast performance, while
\app and anti-affinity-only \appp show only a few well-performing cases---the
rest of the data points at the corresponding performance bracket are shifted to
the right, revealing comparatively slower results. We can characterise the
``fast'' invocations as those where the \textit{divide} and its two
\textit{impera} functions appear on a ``free'' node, i.e., without the
\textit{heavy} function, in Europe. Specifically, when using \app, each
invocation has a $\nicefrac{2}{6}$ probability of appearing on a free node in
Europe, i.e., the probability of fast invocations is $(\nicefrac{2}{6})^3
\approx 3.7\%$, with anti-affinity-only \appp the figure increases to
$(\nicefrac{1}{2})^3 = 12.5\%$ (each invocation has a $\nicefrac{1}{2}$ chance
of appearing on a European free node) and with \appp the probability raises to
50\% since all three functions go on the same node (either in the US or in the
EU).

\begin{figure}[t]
  \caption{Comparison of scheduling times between vanilla, \app-, and \appp-based
  OpenWhisk (avg and st dev are in ms).}
  \center
  \begin{adjustbox}{width=0.46\textwidth}
  \begin{tabular}{|l|c|c|c|c|c|c|c|c|}
  \hline
  & \multicolumn{2}{c|}{\textbf{OpenWhisk}} & \multicolumn{2}{c|}{\textbf{APP}} &
  \multicolumn{2}{c|}{\textbf{aAPP}} \\
  \cline{2-7}
  & avg & st dev & avg & st dev & avg & st dev \\ \hline \textit{hello-world} & 0.68 & 1.16 & 0.73 & 1.25 & 0.8 & 1.27 \\
  \hline
  \textit{long-running} & 0.48 & 0.53 & 0.69 & 0.92 & 0.71 & 1.01 \\
  \hline
  \textit{compute-intens.} & 11.57 & 11.92 & 10.17 & 11.67 & 10.01 & 9.66 \\
  \hline
  \textit{DB-acc., light} & 0.65 & 1.31 & 0.85 & 1.62 & 0.83 & 1.31 \\
  \hline
  \textit{DB-acc., heavy} & 0.44 & 0.69 & 0.91 & 1.25 & 1.04 & 1.7 \\
  \hline
  \textit{external service} & 1.28 & 2.08 & 1.95 & 3.33 & 1.49 & 2.5 \\
  \hline
  \textit{code dependen.} & 0.64 & 1.06 & 1.0 & 2.27 & 0.86 & 1.8 \\
  \hline
  \end{tabular}
  \end{adjustbox}
  
  \label{fig:results_overhead}
\end{figure}

Overall, already introducing anti-affinities improves performance (mean, median,
tail latency improve resp.\@ of 110\%, 10\%, and 178\%), which shows the impact
of sharing a worker with \textit{heavy} functions---\app shows a long tail of
invocations after the ca. 3000ms mark in \cref{fig:benchmark}. Looking at worst
cases, using \appp does not result in a considerable performance increase. This
is visible from the plot by noticing how the tail high-percentage instances of
anti-affinity-only \appp and \appp almost overlap, resulting in a small (+13\%)
improvement in tail latency. The differences in mean (+40\%) and median (+91\%)
latency between having affinities or not emerge in the 250--1000ms bracket,
where the lack of affinity leads to fewer fast executions, in contrast
with the abundance of fast \appp instances. Practically, the figures and
distribution show how strongly North American allocations impact latency vs the
benefit of co-location. Besides increasing performance, \appp succeeds in
eliminating database access retries, contrarily to anti-affinity-only \appp
(i.e., 42 requests suffer at least one retry in \app, 23 in anti-affinity-only
\appp, and 0 in \appp).




%% file: benchmarks.tex
\section{\appp's Overhead is Negligible}
\label{sec:non_affinity_aware}

\begin{figure*}[t]
      \centering
      \includegraphics[width=0.32\textwidth]{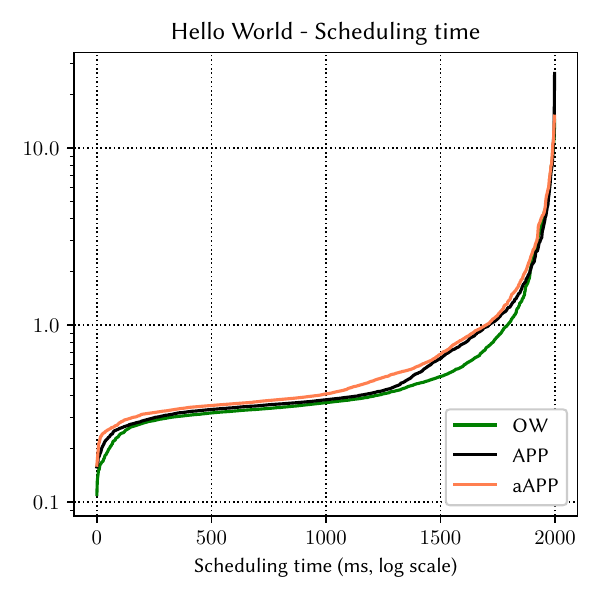}
      \includegraphics[width=0.32\textwidth]{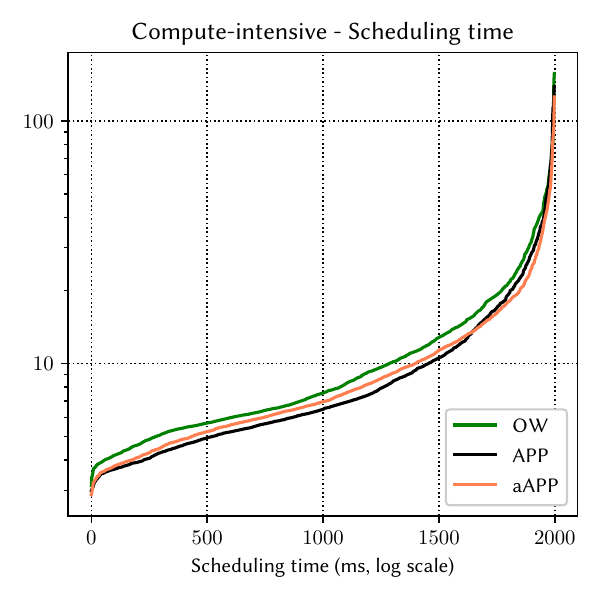}
      \includegraphics[width=0.32\textwidth]{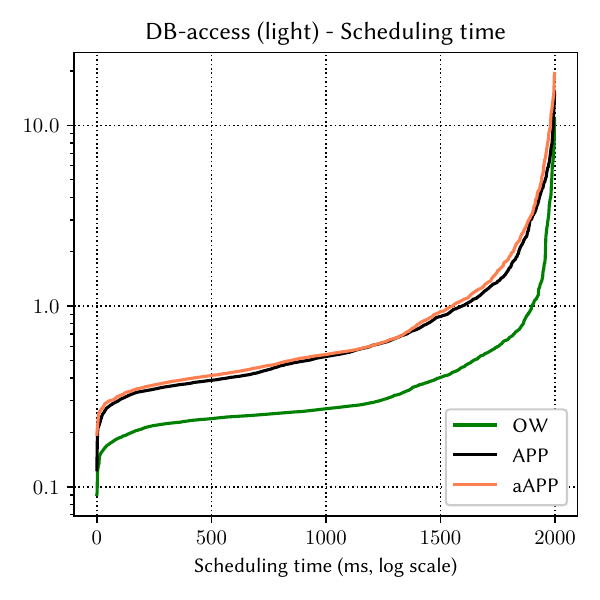}
      \includegraphics[width=0.32\textwidth]{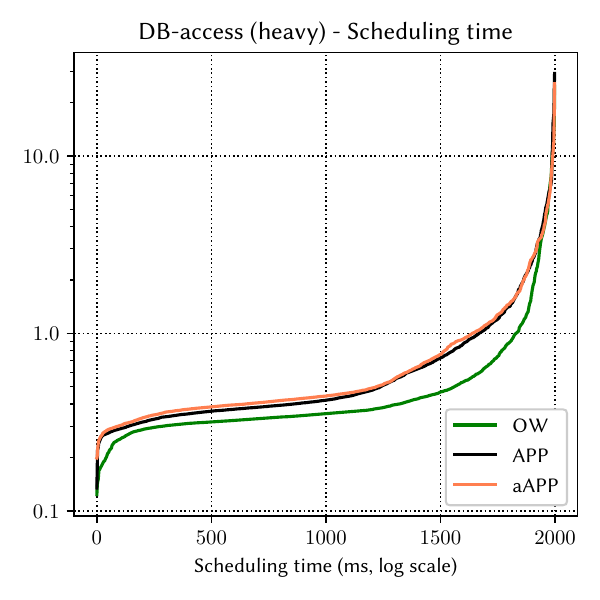}
      \includegraphics[width=0.32\textwidth]{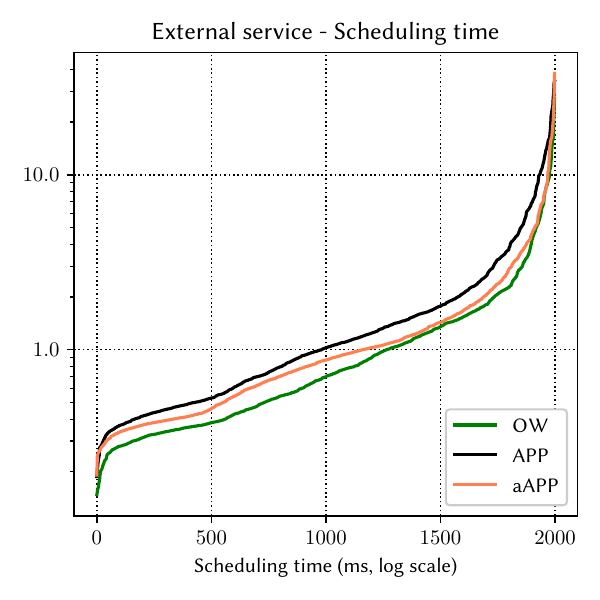}
      \includegraphics[width=0.32\textwidth]{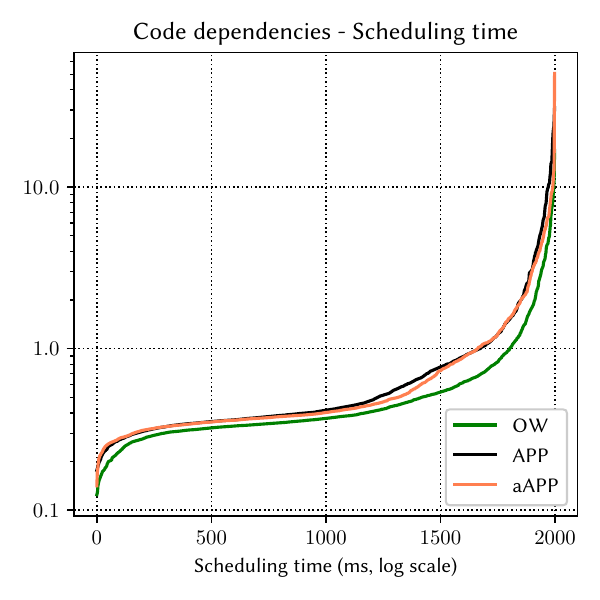}
    
      \caption{Distribution of scheduling times for vanilla, \app-, and \appp-based
      OpenWhisk (x-axis represents instances sorted from the quickest scheduling time to the slowest).}
      \label{fig:function_plots}
\end{figure*}


While \appp allows us to exploit (anti-)affinity constraints that would not be
expressible otherwise, it is crucial to assess the overhead introduced by the
additional functionalities of \appp. In this section, we show that the added
functionalities (to track the state of functions on workers) of our \appp-based
prototype have negligible impact on the platform's scheduling performance.



For these experiments, we use the benchmark suite used by De Palma et
al.~\cite{DGMTZ24} to benchmark their \app-based OpenWhisk implementation. Note
that, in the context of these experiments, we are not interested in the data
locality capabilities of \app, but only in checking the scheduling performances
of \appp. Thus, we deploy the platforms in only one cloud zone and use 2000
invocations for each scenario, to simplify as much as possible the testing
environment and have enough invocations to draw meaningful comparisons.
The benchmarks are:

\begin{itemize}

    \item \textit{hello-world} implements a simple echo application, and
          indicates the baseline performance of the platform;

    \item \textit{long-running} waits for 3 seconds before responding to
          benchmark the handling of multiple functions running for several
          seconds and the management of their queueing process;

    \item \textit{compute-intensive} multiplies two \(10^2\) square matrices and
          returns the result to the caller, measuring both the performance of
          handling functions performing some meaningful computation and of
          handling large invocation payloads;

    \item \textit{DB-access (light)} executes a query for a document from a
          remote MongoDB database. The requested document is lightweight,
          corresponding to a JSON document of 106 bytes, with little impact on
          computation. De Palma et al.\@ used the case to measure the impact of
          data locality on the overall latency. Since we have all workers in the
          same cloud zone, we use it to measure the overhead of scheduling
          functions that fetch small payloads from a local database;

    \item \textit{DB-access (heavy)} regards both a memory- and bandwidth-heavy
          data-query function. The function fetches a large document (124.38 MB)
          from a MongoDB database and extracts a property from the returned
          JSON. Similarly to the previous function, we use this one to evaluate
          the overhead of scheduling functions that fetch large payloads from a
          local database;

    \item \textit{External service} tracks the performance of invoking an
          external API (Slack). De Palma et al.\@ drew the function from the
          Wonderless dataset~\cite{Eskandani-S:Wonderless};

    \item \textit{Code dependencies} is a formatter that takes a JSON string and
          returns a plain-text one, translating the key-value pairings into
          Python-compatible dictionary assignments. De Palma et al.\@ drew
          this case from the Wonderless dataset~\cite{Eskandani-S:Wonderless}.
\end{itemize}

For completeness, we note that we omitted the \textit{cold-start} case from De
Palma et al.~\cite{DGMTZ24}, which is an echo application with sizable, unused
dependencies. The peculiarity of the case is its 10-minute invocation pattern,
used to track cold-start times (so that the platform evicts cached copies of the
function, requiring costly fetch-and-startup times at any subsequent
invocation). We omit this test since we can observe its effects with the
\textit{hello-world} and \textit{code-dependency} cases.

We run the benchmarks on a one-zone Google Cloud cluster with four Ubuntu 20.04
virtual machines with 4 GB RAM each, one with 2 vCPU for the OpenWhisk
controller and three with 1 vCPU, resp.\ for two workers and a MongoDB instance
for the \textit{DB-access} cases.
We run 2000 function invocations for each case in batches of 4 parallel requests
(500 per thread), recording both the scheduling time (the time between the
arrival of a request at the controller and the issuing of the allocation) and
the execution latencies.
We compare \appp, \app, and vanilla OpenWhisk. For a fair comparison 
with vanilla OpenWhisk, we set the \app/\appp configurations with a
\texttt{default} policy that falls back to the vanilla scheduler.

For all cases and platforms, we report in \cref{fig:results_overhead}, in
tabular form, the average (avg) and standard deviation (st dev) of the
scheduling time. On average, all platforms allocate functions in less than 2ms,
except for the \textit{compute-intensive} case, which takes less than 12ms
(likely due to the large request payloads that the controller needs to forward
to workers). As expected, OpenWhisk vanilla is the fastest, closely (under one
millisecond) followed by \app and \appp---except for the
\textit{compute-intensive} case, where \app and \appp perform better and
OpenWhisk is slower by less than 2ms. The differences between \app and \appp are
even smaller, with \app being generally slightly (sub-millisecond) faster than
\appp.
To better characterise the comparison, in \cref{fig:function_plots}, we show the
plot-line distribution of the scheduling times.
The curves exhibit the typical tail distribution pattern~\cite{DB13} of cloud
workloads (which accounts for the high standard deviation reported in
\cref{fig:results_overhead}) and confirm our observations. Excluding the tails,
the two plots almost overlap with negligible, sub-millisecond differences. All
the results reported in \cref{fig:benchmark,fig:function_plots} differ
significantly according to the Wilcoxon test (p = 0.001).

%% file: related_work.tex
\section{Related Work and Conclusion}
\label{sec:conclusion}

To the best of our knowledge, \appp is the first language that allows developers
to state affinity constraints to better control the scheduling of the functions
in FaaS platforms. By extending OpenWhisk, 
we demonstrate the effectiveness of using (anti-)affinity constraint of \appp in
reducing latency and tail latency.
Furthermore, we benchmark that the overhead of supporting \appp-based affinity
constraints is minimal compared to vanilla OpenWhisk and its \app-based variant. 


Broadening our scope, the works we see the closest to ours come from the
neighbouring area of microservices~\cite{DGLMMMS17}---the state-of-the-art style
alternative to serverless for cloud architectures. Proposals in this direction
are by Baarzi and George~\cite{BK21}, who present a framework for the deployment
of microservices that infers and assigns affinity and anti-affinity traits to
microservices to orient the distribution of resources and microservices replicas
on the available machines; Sampaio et al.~\cite{SRBR19}, who introduce an
adaptation mechanism for microservice deployment based on microservice
affinities (e.g., the more messages microservices exchange the more affine they
are) and resource usage; Sheoran et al.\cite{SFSM21}, who propose an approach
that computes procedural affinity of communication among microservices to make
placement decisions. Looking at the industry, Azure Service
Fabric~\cite{web:azure_sf} provides a notion of \emph{service affinity} that
ensures that the replicas of a service are placed on the same nodes as those of
another, affine service. Another example is Kubernetes, which has a notion of
\emph{node affinity} and \emph{inter-pod (anti-)affinity} to express advanced
scheduling logic~\cite{web:kube_affinity}.

Overall, the mentioned work proves the usefulness of affinity-aware deployments
at lower layers than FaaS (e.g., VMs, containers, microservices) and compels a
discussion on the interplay between \appp and IaaS/CaaS-level affinity, which we
detail under two main directions.
On the one hand, one could realise a version of \appp for the Infrastructure
and/or the Container layers. We argue it is more interesting to focus on FaaS.
Indeed, there are mainstream IaaS and CaaS platforms that allow users to program
directly ad-hoc schedulers (e.g., Kubernetes exposes APIs for creating scheduler
plugins that define its scheduling policies). Since these layers afford a higher
level of customisation than \appp---at the expense of more technical involvement
on the part of the users---a variant of \appp for the IaaS/CaaS-levels seems
less useful.
On the other hand, one can use IaaS and CaaS platforms that support affinity
constraints to implement affinity-aware FaaS platforms. We see two main problems
with pursuing this path. The first regards performance. To implement FaaS-level
affinity using IaaS/CaaS affinity constraints, we need to impose a 1:1 relation
between a function instance and the VM/container running it (if we let the same
VM/container run parallel copies of the same function we cannot guarantee e.g.,
self anti-affinity). This imposition would prevent the platform from exploiting
the serverless technique of VM/container reuse to avoid cold
starts~\cite{Mohan-etal:AgileColdStartsForScalableServerless,SWLSGHT20,SA20}.
The second problem regards abstraction leakage, where letting FaaS users access
the underlying IaaS/CaaS layers leaks details 
of the infrastructural components breaking FaaS' paradigmatic abstractions.


%

A recent trend of FaaS is the definition/handling of the composition/workflows
of functions, like AWS step-functions~\cite{web:step_functions} and Azure
Durable functions~\cite{BGJKMM21}. The main idea behind these works is to allow
users to define workflows as the composition of functions with their branching
logic, parallel execution, and error handling. The orchestrator/controller of
the platform then uses the workflow to manage function executions and handle
retries, timeouts, and errors. Our proposal is orthogonal to these works.
Indeed, assuming the workflow is available, the orchestrator developed for
handling serverless workflows should be extensible with an \appp-like script to
specify where to schedule the functions within a given workflow. Future work on
this integration would support the enforcement of even more expressive policies
than \appp, like preventing function instances of the same workflow from sharing
nodes.

%
Another interesting proposal, Palette~\cite{AGLFCGBBF23}, uses optional opaque
parameters in function invocations to inform the load balancer of Azure
Functions on the affinity with previous invocations and the data they produced.
While Palette does not support (anti-)affinity constraints, it allows users to
express which invocations benefit from running on the same node. We deem an
interesting future work extending \appp to infer (anti-)affinity constraints based on observed performance of different deployments or by
inspecting the functions code.
We already started exploring this landscape by
introducing new options in \app{} that use static analysis and dynamic
monitoring to select the workers that minimise the latency of calling external
services in functions~\cite{DGLMTZ23}.

Regarding the constructs we have proposed for expressing affinity-aware policies
in \appp, we observe that an alternative approach could be to let the user
directly declare the properties to enforce, leaving to the platform the task to
realise them at run time.
The scheduling runtime of this \app variant would allocate a function only if
the allocation satisfies the formula or fail otherwise. The limitation of this
approach lies in its scalability. Verifying the satisfiability of a property
could require assessing multiple interacting constraints, possibly leading to an
exponential time complexity with respect to the formula's size, the number of
workers, and the number of functions involved.\footnote{For example, if the
language allows encoding properties such as ``schedule the function $f$ only if
it does not prevent scheduling higher priority functions $g_1, \dots, g_n$''
then, due to the NP-hardness of bin-packing~\cite{DBLP:books/fm/GareyJ79}, the
problem of scheduling a function becomes an NP-hard problem.} Contrarily, the
\appp scheduler checks whether it can allocate a function on a worker in linear
time on the size of the workers and \appp script length.

While, in this work, we focus on the design and run time performances 
of (anti-)affinity scheduling policies, we have also investigated the 
problem of statically
checking scheduling properties of \app{} and \appp{} scripts~\cite{DGMTZ23}.
For instance, given the scheduling policy expressed by a script,
one could be interested in checking the possibility for one safety critical
function to be scheduled on an untrusted worker, or the possibility for 
the same function to be contemporaneously scheduled on the same worker
with an ``unknown'' function developed by a third party.
The approach we have investigated consists of the translation of the 
reachability property of interest into a planning problem, and then exploit 
off-the-shelf planners to solve such problem.




An important future research direction is empirically investigating the
\appp{}'s usability and intuitiveness by conducting experimental studies that
probe developers' ability to understand and specify (anti-)affinity constraints
to implement properties of function scheduling (e.g., that two given functions
are never scheduled on the same worker). Such investigations would both serve to
validate the language's design and uncover potential implicit understanding or
unexpected interpretation patterns that might emerge when developers engage with
\appp{} with different levels of preconditioning.

Implementation-wise, we are planning to extend the support for \appp{} to other serverless platforms. We already started this work by extending the support for a variant of \app{} in the FunLess serverless platform~\cite{DGLMTZ25}. Natural steps in this direction include integrating \app{} with other popular FaaS platforms like Knative and OpenFaaS.

Regarding the integration within OpenWhisk, the platform supports scenarios
where multiple controllers share the pool of available workers (e.g., for
redundancy and load balancing) and take scheduling decisions without
coordination. In our \appp-based implementation, such multi-controller
configurations present a problem since we need to prevent scheduling races among
controllers---e.g., imagine two controllers that select an available, empty
worker and, at the same time, allocate mutually anti-affine functions on it.
Supporting multi-controller deployments is outside the scope of this paper and an interesting subject for future
work.

Finally, while our evaluation demonstrates that affinity and anti-affinity
constraints can enhance serverless application performance, a standardised
benchmark of real-world examples would enable a more comprehensive analysis.
Unfortunately, to the best of our knowledge, no such benchmark currently exists.
As a direction for future work, we
plan to collaborate with the community to collect instances of real-world
applications and their affinity constraints, creating a benchmark to compare
scheduler performances in serverless platforms.